\documentclass[preprintnumbers,amsmath,amssymb,nofootinbib]{revtex4}
\usepackage{graphicx}% Include figure files
\usepackage{epsfig}
\usepackage[dvips]{color}

\begin{document}

\begin{center}
{\Large {\bf Analysis on Heavy Quarkonia Transitions with Pion
Emission in Terms of the QCD Multipole Expansion and Determination
of Mass Spectra of Hybrids }}
\end{center}

\vspace*{0.3cm}
\begin{center}
{Hong-Wei Ke, Jian Tang, Xi-Qing Hao and Xue-Qian
Li  }\\
\vspace*{0.3cm}
{\it Department of Physics, Nankai University, Tianjin 300071, China}\\

\end{center}

\vspace*{0.5cm}

\begin{center}
\begin{minipage}{12cm}

\noindent Abstract:\\
One of the most important tasks in high energy physics is search
for the exotic states, such as glueball, hybrid and multi-quark
states. The transitions $\psi(ns)\rightarrow \psi(ms)+\pi\pi$ and
$\Upsilon(ns)\rightarrow \Upsilon(ms)+\pi\pi$ attract great
attentions because they may reveal characteristics of hybrids. In
this work, we analyze those transition modes in terms of the
theoretical framework established by Yan and Kuang. It is
interesting to notice that the intermediate states between the two
gluon-emissions are hybrids, therefore by fitting the data, we are
able to determine the mass spectra of hybrids. The ground hybrid
states are predicted as 4.23 GeV (for charmonium) and 10.79 GeV
(for bottonium) which  do not correspond to any states measured in
recent experiments, thus it may imply that very possibly, hybrids
mix with regular quarkonia to constitute physical states.
Comprehensive comparisons of the potentials for hybrids whose
parameters are obtained in this scenario with the lattice results
are presented. \noindent \vspace*{0.5cm}

\end{minipage}

\end{center}

\baselineskip 22pt
\newpage

\section{Introduction}

In both the quark model and QCD which governs strong interaction,
there is no any fundamental principle to prohibit existence of
exotic hadron states such as glueball, hybrid and multi-quark
states. In fact, to eventually understand the low energy behavior
of QCD, one needs to find out such states. However, the recent
research indicates that they may mix with the ordinary hadrons
especially the quarkonia. Thus they evade direct detection so far,
even though many new resonances which have peculiar
characteristics, have continuously been reported by various
experimental collaborations. Theorists have proposed them to be
pure gluonic (glueball), quark-gluon (hybrid), and/or multi-quark
(tetraqurk or pentaquark) structures which are different from the
regular valence quark structure of $q\bar q$ for meson and $qqq$
for baryon. Since the quark model and QCD theory advocate their
existence, at least do not repel them, one should find them in
experiments. However, even with many candidates of the exotic
states, so far none of them have been confirmed yet. Moreover, the
possible mixing of such exotic states with the regular mesons or
baryons contaminates the situation and would make a clear
identification difficult, even though not impossible. From the
theoretic aspect, one may try to help to clean the mist and find
an effective way to do the job.

The transition of heavy quarkonia such as $\psi(ns)$ and $\Upsilon
(ns)$ to lower states $\psi(ms)$ and $\Upsilon(ms)$  ($m<n$) with
two pions being emitted, provides an ideal laboratory to study the
spectra of hybrids. In the transitions
$\psi(ns)(\Upsilon(ns))\rightarrow \psi(ms)(\Upsilon(ms))+\pi\pi$
($m<n$), the momentum transfer is not large and usually the
perturbative method does not apply. The QCD multipole expansion
(QCDME) method suggested by Gottfried, Yan and
Kuang\cite{Gottfried,YK1,K2,Y1,K1} well solves the light-meson
emission problem. In the picture of the multipole expansion, two
gluons are emitted which are not described as energetic particles,
but a chromo filed of TM or TE modes, then the two gluons which
constitute a color singlet, hadronize into light
hadrons\cite{soft}. It is worth emphasizing again that the two
gluons are not free gluons in the sense of the perturbative
quantum field theory, but a field in the QCD multipole expansion.
It is easy to understand that such transition is dominated by the
E1-E1 mode, while the M1-M1 mode is suppressed for the heavy
quarkonia case.

Since two gluons are successively emitted, there exists an
intermediate state where the quark-antiquark pair resides in a
color octet. The color octet $q-\bar q$ and a color-octet gluon
constitute a color singlet hybrid state. Therefore, in the
framework, a key point is to determine the spectra of the hybrid
states $|q\bar qg>$ where $q$ can be either $b$ or $c$ in our
case. Due to lack of enough data to fix the ground state of hybrid
mesons, Buchm\"uller and Tye \cite{BT} assumed that the observed
$\psi(4.03)$ was the ground state of $|c\bar cg>$.

Yan and Kuang used this postulate to carry out their estimation on
the transition rates\cite{YK1,K2}. For the intermediate hybrid
states they used the phenomenological potential given by
Buchm\"uller and Tye\cite{BT} to calculate the widths of
$\Upsilon(2s)\rightarrow \Upsilon(1s)\pi\pi$,
$\Upsilon(3s)\rightarrow
\Upsilon(1s)\pi\pi$,$\Upsilon(3s)\rightarrow \Upsilon(2s)\pi\pi$.
The theoretical prediction on the rate of $\Upsilon(2s)\rightarrow
\Upsilon(1s)\pi\pi$ and $\Upsilon(3s)\rightarrow
\Upsilon(2s)\pi\pi$ is roughly consistent with data\cite{PDG},
whereas that for $\Upsilon(3s)\rightarrow \Upsilon(1s)\pi\pi$
obviously deviates from data. It is also noted that when they
calculated the decay widths, they need to invoke a cancellation
among large numbers to obtain smaller physical quantities, thus
the calculations are very sensitive to the model parameters, i.e.
a fine-tuning is unavoidable. Recently Kuang \cite{K2} indicates
that determining the proper intermediate hybrid states is crucial
to predict the rates of the decay modes such as
$\Upsilon(3s)\rightarrow \Upsilon(1s)\pi\pi$.

There have been some models for evaluating the hybrid spectra, but
there are several free parameters in each model and one should
determine them by fitting data. This leads to an embarrassing
situation that one has to determine at least one hybrid state, and
then obtain the corresponding parameters in the model. Moreover,
the recent studies indicate that hybrid may not exist as an
independent physical state, but mixes with regular quarkonia
states, therefore the mass spectra listed on the data table are
not the masses of a pure hybrid, which are the eigenvalues of the
Hamiltonian matrices. Therefore a crucial task is to determine the
mass spectra of pure hybrids, even though they are not physical
eigenstates of the Hamiltonian matrices.

Recently, thanks to the progress of measurements of the Babar
\cite{4s2s} and Belle \cite{Belle} collaborations, a remarkable
amount of data on the transitions
$\psi(ns)(\Upsilon(ns))\rightarrow \psi(ms)(\Upsilon(ms))+\pi\pi$
have been accumulated and become more accurate. Since the large
database is available, one may have a chance to use the data to
determine the mass spectra of hybrids.

In this work, we apply the QCD multipole expansion method
established by Yan and Kuang \cite{YK1} and the potential model
given by several groups \cite{Isgur,Swanson,Allen}, to calculate
the transition rates of $\psi(ns)(\Upsilon(ns))\rightarrow
\psi(ms)(\Upsilon(ms))+\pi\pi$ by keeping the potential model
parameters free. Then by the typical method, namely minimizing
$\bar{\chi}^2$ for the channels which have been well measured, we
obtain the corresponding parameters, and then we go on predicting
a few channels which are not been measured yet, finally with the
potential we can determine the masses of  hybrids, at least the
ground state.

To make sense, we compare the potentials for hybrids whose
parameters are obtained in this scenario with the results of the
lattice calculation. We find that if the parameters in the
potential suggested by Allen et al.\cite{Allen} adopt the values
which are obtained in terms of our strategy, the potential
satisfactorily coincides with the lattice results.

Our numerical results indicate that the ground states of pure
hybrid $|c\bar cg>$ and $|b\bar bg>$ do not correspond to the
physical states measured in recent experiments, the concrete
numbers may somehow depend on the forms of the potential model
adopted for the calculations (see the text). This may suggest that
the pure hybrids do not exist independently, but mix with regular
mesons.

After the introduction we present all the formulation in next
section, where we only keep the necessary expressions for later
calculations, but omitting some details which can be easily found
in Yan and Kuang's papers. Then we carry out our numerical
analysis in term of the $\bar{\chi}^2$ method. Comprehensive
comparisons of various potentials  with the lattice results are
presented. The last section is devoted to conclusion and
discussion.

\begin{figure}[!h]\label{p1}
\begin{center}
%\begin{tabular}{ccc}
\scalebox{0.8}{\includegraphics{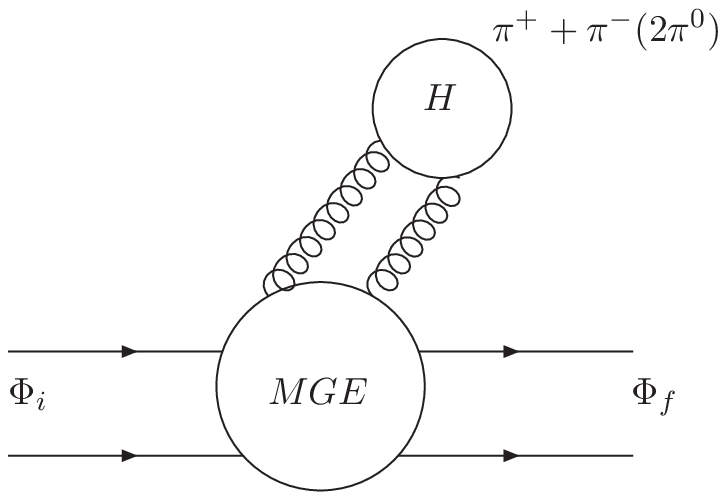}}
%\end{tabular}
\end{center}
\caption{}
\end{figure}

\section{Formulation}
\subsection{The transition width }
The theoretical framework about the QCD Multiploe Expandsion
method is well established in Refs\cite{YK1,K2,Y1,K1}, and all the
corresponding formulas are presented in their series of papers.
Here we only make a brief introduction to the formulas for
evaluating the widths which we are going to employ in this work.
In Refs.\cite{YK1,K2} the transition rate of a vector quarkonium
into another vector quarkonium with a two-pion emission can be
written as
\begin{eqnarray}\label{s1}
\Gamma(n_{I}{}^{3}S_{1}\rightarrow
n_{F}{}^{3}S_{1})=|C_1|^2G|f^{l,P_{I},P_{F}}_{n_{I},l_{I},n_{F},l_{F}}|^2
\end{eqnarray}
where $|C_1|^2$ is a constant to be determined and it comes from
the hadronization of gluons into pions, $G$ is the phase space
factor, $f^{l,P_{I},P_{F}}_{n_{I},l_{I},n_{F},l_{F}}$ is the
overlapping integration over the concerned hadronic wave
functions, their concrete forms were given in \cite{K2} as
\begin{eqnarray}\label{s2}
f^{l,P_{I},P_{F}}_{n_{I},l_{I},n_{F},l_{F}}=\sum_K\frac{\int
R_F(r)r^{P_F}R^*_{Kl}(r)r^2dr \int
R^*_{Kl}(r')r'^{P_I}R_I(r')r'^2dr'}{M_I-E_{Kl}},
\end{eqnarray}
where $n_{I},n_{F}$ are the principal quantum numbers of initial
and final states, $l_{I},l_{F}$ are the angular momenta of the
initial and final states, $l$ is the angular momentum of the
color-octet $q\bar{q}$ in the intermediate state, $P_{I},P_{F}$
are the indices related to the multipole radiation, for the E1
radiation $P_{I},P_{F}$=1 and $l=1$. $R_I,R_F$ and $R_{Kl}$ are
the radial wave functions of the initial and final states, $M_I$
is the mass of initial quarkonium and $E_{Kl}$ is the energy
eigenvalue of the intermediate hybrid state.

\subsection{The $\bar{\chi}^2$ method}

The standard method adopted in analyzing data and extracting useful
information is minimizing the $\bar{\chi}^2$ and in our work, we
hope to obtain the model parameters. When calculating
$\bar{\chi}^2$, we would involve as many as possible experimental
measurements to make the fitted parameters more reasonable. Here we
adopt the form of $\bar{\chi}^2$ defined in \cite{chi} as
\begin{eqnarray}\label{ks}
\bar{\chi}^2=\sum_i\frac{(W^{th}_i-W^{exp}_i)^2}{(\Delta
W^{exp}_i)^2},
\end{eqnarray}
where $i$ represents the i-th channel, $W^{th}_i$ is the
theoretical prediction on the width of channel $i$, $W^{exp}_i$ is
the corresponding experimentally measured value, $\Delta
W^{exp}_i$ is the experimental error.

$W^{th}_i$ will be calculated in terms of the potential models
with several free parameters which are described in the following
subsections, thus $W^{th}_i$ is a function of the parameters. By
minimizing $\bar\chi^2$, we would expect to determine the model
parameters. Some details of our strategy will be depicted in
subsection E.

\subsection{The phenomenological potential for the initial and final quarkonia}

In this work, we adopt two different potentials for the initial
and final heavy quarkonia and the intermediate hybrid states.

The Cornell potential \cite{cornell} is the most popular potential
form to study heavy quarkonia.  The potential reads as
\begin{eqnarray}\label{cornell}
V(r)=-\frac{\kappa}{r}+br,
\end{eqnarray}
usually in the literature many authors prefer to use $\alpha_s$
instead of $\kappa$ and it has a relation $
\kappa=\frac{4\alpha_s(r)}{3}$, and $\alpha_s(r)$ can be treated
as a constant for the $\bar bb$ and $\bar cc$ quarkonia.

The modifed Cornell potential: It may be more reasonable to choose
a modified Cornell potential which includes a spin-related term
\cite{ss}, and the potential takes the form
\begin{eqnarray}\label{s5}
V(r)=-\frac{\kappa}{r}+br+V_{s}(r)+V_0,
\end{eqnarray}
where the spin-related term $V_{s}$ is,
\begin{eqnarray*}
&&V_{s}=\frac{8\pi\kappa}{3m_q^2}\delta_\sigma(r)\overrightarrow{S}_q
\cdot\overrightarrow{S}_{\bar{q}}
\end{eqnarray*}
with
\begin{eqnarray*}
\delta_\sigma(r)=(\frac{\sigma}{\sqrt{\pi}})^3e^{-\sigma^2r^2},
\end{eqnarray*}
and $V_{0}$ is the zero-point energy,( in Ref.\cite{ss} it was set
to be zero), here we do not priori-assume it to be zero, but fix
it by fitting the spectra of heavy quarkonia.

\subsection{The potential for hybrids}

The intermediate state as discussed above is a hybrid state
$|q\bar qg>$ and we need to obtain the spectra and wave-functions
of the ground state and corresponding radially excited states. Yan
and Kuang used the phenomenological potential given by
Buchm\"uller and Tye \cite{BT} to evaluate the mass of the ground
state of hybrid, instead, in our work, we take some effective
potential models which are based on the color-flux-tube model.

Generally hybrids are labelled by the right-handed($n_m^+$) and
left-handed($n_m^-$) transverse phonon modes
\begin{eqnarray*}\label{h1}
N=\sum_{m=1}^\infty m(n_m^++n_m^-),
\end{eqnarray*}
and a characteristic quantity $\Lambda$ as
 \begin{eqnarray*}\label{h2}
\Lambda=\sum_{m=1}^\infty (n^+_{m}-n^-_{m}).
\end{eqnarray*}

All the details about the definitions and notations can be easily
found in
literature\cite{Isgur,Swanson,Allen,Buisseret,Szczepaniak,Bali}.

Various groups suggested different potential forms for the
interaction between the quark and antiquark in the hybrid state.
We label them as Model 1, 2 and 3 respectively.

In this work, we employ three potentials which are:

Model 1 was suggested by Isgur and Paton \cite{Isgur} as
\begin{eqnarray}\label{h4}
V(r)=-\frac{\kappa}{r}+br+\frac{\pi}{r}(1-e^{-fb^{1/2}r})+V_0.
\end{eqnarray}

Model 2: Swanson and Szczepaniak\cite{Swanson} think that the
Coulomb term in model 1 is not compatible with the lattice
results, so that they suggested an alternative effective potential
as
\begin{eqnarray}\label{h5}
V(r)=br+\frac{\pi}{r}(1-e^{-fb^{1/2}r}).
\end{eqnarray}

To get a better fit to data, we add the zero-point energy $V_0$
into Eq.(\ref{h5}),
\begin{eqnarray}\label{h6}
V(r)=br+\frac{\pi}{r}(1-e^{-fb^{1/2}r})+V_0.
\end{eqnarray}

Model 3: In model 1, the Coulomb piece is not proper, because the
quark and antiquark in the hybrid reside in a color-octet instead
of a singlet (the meson case), the short-distance behavior should
be repulsive (it is determined by the sign of the expectation
value of the Casimir operator in octet). Thus Allen $et\, al$
suggested the third model \cite{Allen} and the corresponding
potential form is
\begin{eqnarray}\label{h7}
V(r)=\frac{\kappa}{8}+\sqrt{(br)^2+2\pi b}+V_0.
\end{eqnarray}

Because in these forms the authors do not consider the
spin-related term (which we name as $V_s$.), we can modify the
potential by adding a spin-related term $V_s$, then the potential
becomes:
\begin{eqnarray}\label{s91}
V(r)=V_i+V_s.
\end{eqnarray}
By this modification, one can investigate the spin-splitting
effects. Generally, $V_s$ should have the same form as that in
(\ref{s5}).

\subsection{Our strategy}

The strategy of this work is that we will determine the concerned
parameters in the potential (Eqs.(\ref{h4}), (\ref{h6}),
(\ref{h7}) and Eq.(\ref{s91})) by fitting the data of heavy
quarkonia transitions.

To obtain the concerned parameters in the potentials
(Eqs.(\ref{h4}), (\ref{h6}), (\ref{h7}) and (\ref{s91})) which
specify the hybrids sates,  we use the method of minimizing
$\bar\chi^2$ defined in (\ref{ks}). Concretely, in Eq.(\ref{ks}),
$W^{th}_i$ is a function of the parameters $\kappa,\; f,\;
b,\;V_0$ and $|C_1|^2$, and following Ref.\cite{Isgur}, we set
$f=1$, therefore $\bar \chi^2$ is also a function of those
parameters. Minimizing $\bar\chi^2$, one can fix the values of the
corresponding parameters. Still for simplifying our complicated
numerical computations, we choose a special method, namely, we
first pre-set a group of the parameters, and we calculate the
hybrid spectra and wave-functions by solving the Schr\"odinger
equation, then we determine $|C_1|^2$ in Eq.(\ref{s1}) in terms of
the well measured rate of $\psi(2S)\rightarrow J/\psi\,\pi\,\pi$.
With this $|C_1|^2$ as a pre-determined value or say, a function
of other parameters, we minimize $\bar\chi^2$ to fix the values of
the rest of parameters $\kappa,\; b,\;V_0$.

With all the parameters being fixed, we can determine the mass
spectra of the hybrids which serve as the intermediate states in
the transitions of $\psi(ns)(\Upsilon(ns))\rightarrow
\psi(ms)(\Upsilon(ms))+\pi\pi$. It is noted that the spectra
determined in this scheme are not really the masses of physical
states, unless the hybrids do not mix with regular quarkonia. In
other words, we would determine a diagonal element of the mass
Hamiltonian matrix, whose diagonalization would mix the hybrid and
quarkonium and then determine the eigenvalues and eigen-functions
corresponding to the physical masses and physical states which are
measured in experiments.

\section{Numerical results}

To determine the model parameters in the potential, we need to fit
the spectra of $\psi(ns)$, $\eta_c(1s)$, $\eta_c(2s)$ and
$\Upsilon(ns)$ and in this work, we only concern the ground states
and radially excited states of $c\bar c$, $b\bar b$ and $c\bar
cg$, $b\bar bg$ systems.

\subsection{ Without the spin-related term $V_s$}

The potentials for quarkonia (Eq.(\ref{cornell})) and hybrid
(Eq.(\ref{h4}) (model 1), Eq.(\ref{h6}) (model 2) and
Eq.(\ref{h7}) (model 3)) do not include the spin-related term. In
this work, we adopt the Cornell potential to calculate the spectra
and wavefunctions of the regular heavy quarkonia. The concerned
parameters in the Cornell potential have been given in literature
as for the $c\bar{c}$ mesons, $\kappa=0.52,\;b=0.18 \rm{GeV^2},\;
m_c=1.84$ GeV, whereas for the $b\bar{b}$ mesons, $\kappa=0.48,\;
b=0.18 \rm{GeV^2},\; m_b=5.17$ GeV\cite{YK1,cornell}. It is also
noted that to meet the measured spectra of charmonia and bottonia
a zero-point energy $V_0$ is needed.

The potential for the hybrid takes three possible forms which are
shown in Eq.(\ref{h4}), (\ref{h6}) and (\ref{h7}). We keep the
values $m_c=1.84$ GeV, $m_b=5.17$ GeV which are obtained by
fitting the spectra of regular quarkonia $|b\bar b>\;
(\Upsilon(ns))$ and $|c\bar c>\; (\psi(ns))$ with the potential
(\ref{cornell}), but need to gain the values of the relevant
parameters $\kappa$, b and $V_0$  etc. by minimizing $\bar\chi^2$
for the decays
$\psi(ns)(\Upsilon(ns))\to\psi(ms)(\Upsilon(ms))+\pi\pi$.
According to the measured value for $\Gamma(\psi(2S)\rightarrow
J/\psi\,\pi\,\pi)$:
\begin{eqnarray*}\label{s9}
&&\Gamma_{tot}(\psi(2S))=337\pm13 \rm{keV} \\&&
B(\psi(2S)\rightarrow J/\psi\pi^+\pi^-)=(31.8\pm0.6)\% \\&&
B(\psi(2S)\rightarrow J/\psi\pi^0\pi^0)=(16.46\pm0.35)\%
\end{eqnarray*}
we express $C_1^2$ as a function of the potential parameters which
exist in the three potentials ( Eqs.(\ref{h4}), (\ref{h6}) or
(\ref{h7})) and will be determined. It is noted that $C_1^2$ is a
factor related to the hadronization of gluons into two pions, so
should be universal for both $\psi$ and $\Upsilon$ decays. The
parameters in the potentials are also universal for the $\bar bb$
and $\bar cc$ cases except the masses are different.

Then for $\Gamma(\Upsilon(nS)\to \Upsilon(ms)+\pi\pi)$ ($m<n$), we
calculate $W^{th}_i$ in terms of the three potential forms. The
corresponding experimental values and errors are $W^{exp}_i$ and
$\Delta W^{exp}_i$ given in the references which are shown in
Table \ref{t1}.
\begin{table}[!h]
\caption{transition rate of $\Upsilon(nS)\to\Upsilon(ms)+\pi\pi$,
(in unit of keV)}
\begin{center}
\begin{tabular}{|c|c|c|c|c|c|c|c|}
  \hline
 decay mode & Model 1 &Model 2 & Model 3& Experiment data \\\hline
$\Upsilon(2S)\rightarrow\Upsilon(1S)\pi\pi$&9.36 &9.28&8.69 &$12.0\pm1.8$ \\
$\Upsilon(3S)\rightarrow\Upsilon(1S)\pi\pi$&1.81 &1.67&1.85 &$1.72\pm0.35$ \\
$ \Upsilon(3S)\rightarrow\Upsilon(2S)\pi\pi$&0.86 &0.76&0.86 &$1.26\pm0.40$ \\
$\Upsilon(4S)\rightarrow\Upsilon(1S)\pi\pi$&3.87 &3.43&4.14   &
$3.7\pm0.6\pm0.7$ \cite{Belle}
\\
$\Upsilon(4S)\rightarrow\Upsilon(2S)\pi\pi$&1.83 &0.2&1.44 &$2.7\pm0.8$ \cite{4s2s} \\
  \hline
\end{tabular}\label{t1}
\end{center}
\end{table}

By minimizing $\bar{\chi}^2$ (eq.(\ref{ks})), we finally get the
potential parameters $\kappa,\; b$ and $V_0$ and the resultant
$\bar{\chi}^2$=4.42 for model 1, 13.69 for model 2 and  7.26 for
model 3 . Then we obtain $|C_1|^2= 100.39\times10^{-6}$ for model
1, $259.24\times10^{-6}$ for model 2, and $121.78\times10^{-6}$
for model 3, the other parameters are listed in the following
table (Table\ref{t2}).
\begin{table}[!h]
\caption{ potential parameters for hybrid}
\begin{center}
\begin{tabular}{|c|c|c|c|c|c|}
  \hline
$$ & $\kappa$ &$b$(GeV$^2$) &$V_0$(GeV)\\\hline
  Model 1 & 0.43& 0.19 & -0.85 \\\hline
 Model 2&-& 0.15 & -0.43 \\\hline
   Model 3& 0.59& 0.19 & -0.85 \\\hline
\end{tabular}\label{t2}
\end{center}
\end{table}

With these potential parameters, we solve the Schr\"odinger
equation to obtain  the masses of ground hybrid states of $|c\bar
cg>$ and  $|b\bar bg>$ (Table \ref{t3}). It is noted that the
resultant spectra depend on the potential forms. We will discuss
this problem in the last section.

\begin{table}[!h]
\caption{the mass of hybrids(in units of GeV)}
\begin{center}
\begin{tabular}{|c|c|c|c|c|c|}
  \hline
  & Model 1& Model 2&Model 3\\\hline
   $|c\bar{c}g>$& 4.099 &4.549&4.226 \\\hline
  $|b\bar{b}g>$& 10.560&11.137&10.789  \\
     \hline
\end{tabular}\label{t3}
\end{center}
\end{table}

We also make a prediction on the rates which have not been
measured yet(Table IV).
\begin{table}[!h]
\caption{prediction(in units of KeV)}
\begin{center}
\begin{tabular}{|c|c|c|c|c|c|}
  \hline
 decay mode & Model 1& Model 2 & Model 3\\\hline
   $\Upsilon(4S)\rightarrow \Upsilon(3S)\pi\pi$& 0.60 &0.57&0.61 \\
  $\psi(3S)\rightarrow \psi(2S)\pi\pi$& 14.96&14.45&14.83  \\
   $\psi(3S)\rightarrow \psi(1S)\pi\pi$&589.91 &72.34& 424.22 \\
   \hline
\end{tabular}\label{t4}
\end{center}
\end{table}

It is noted that values of $\Gamma(\psi(3S)\rightarrow
\psi(1S)\pi\pi)$ predicted by models 1, 2 and 3 are quite apart,
while  $\Gamma(\Upsilon(4S)\rightarrow \Upsilon(3S)\pi\pi)$ and
$\Gamma(\psi(3S)\rightarrow \psi(2S)\pi\pi)$ predicted by all the
three models are close.

\subsection{ Comparison with the lattice results}

To make sense, it would be helpful to compare the results obtained
in our phenomenological work with the lattice results which are
supposed to include both perturbative and non-perturbative QCD
effects. Below we show comprehensive comparisons of our potentials
with the lattice results.

Following Refs.\cite{Swanson,Allen,Buisseret,Szczepaniak,lattice},
the potentials shown in Fig.2 are specially scaled by
$V_{\Sigma_g^+}(2r_0)$ which is the potential for $\Sigma_g^+$
(N=0) at $2r_0=5\; {\rm GeV}^{-1}$ (for the vertical axis of
Fig.2.).

In the three graphs of Fig. 2, we present comparisons of the three
potentials (models, 1,2 and 3) with the parameters fixed in last
subsections with the lattice results. In the graphs, the dots are
the lattice values \cite{lattice}.

It is emphasized that we obtain the potential by minimizing
$\bar\chi^2$ of the data on $\psi(ns)(\Upsilon(ns))\to \psi(ms)
(\Upsilon(ms))+\pi\pi$, but do not fit the lattice values. Then
our results, especially the third potential coincides with the
lattice results extremely well. It may indicate that the physics
description adopted in this scenario is reasonable. It is also
noted that by model 1, the short-distance behavior of the
potential is attractive and obviously distinct from the lattice
results. This discrepancy was discussed above that the
quark-antiquark system in hybrid should be a color-octet and
short-distance interaction should be repulsive. The second
potential (model 2) have the same trend as the lattice results,
but have obvious deviations (see the graph 2 of Fig. 2).

\begin{figure}[!h]\label{p2}
\begin{center}
\begin{tabular}{ccc}
\scalebox{0.8}{\includegraphics{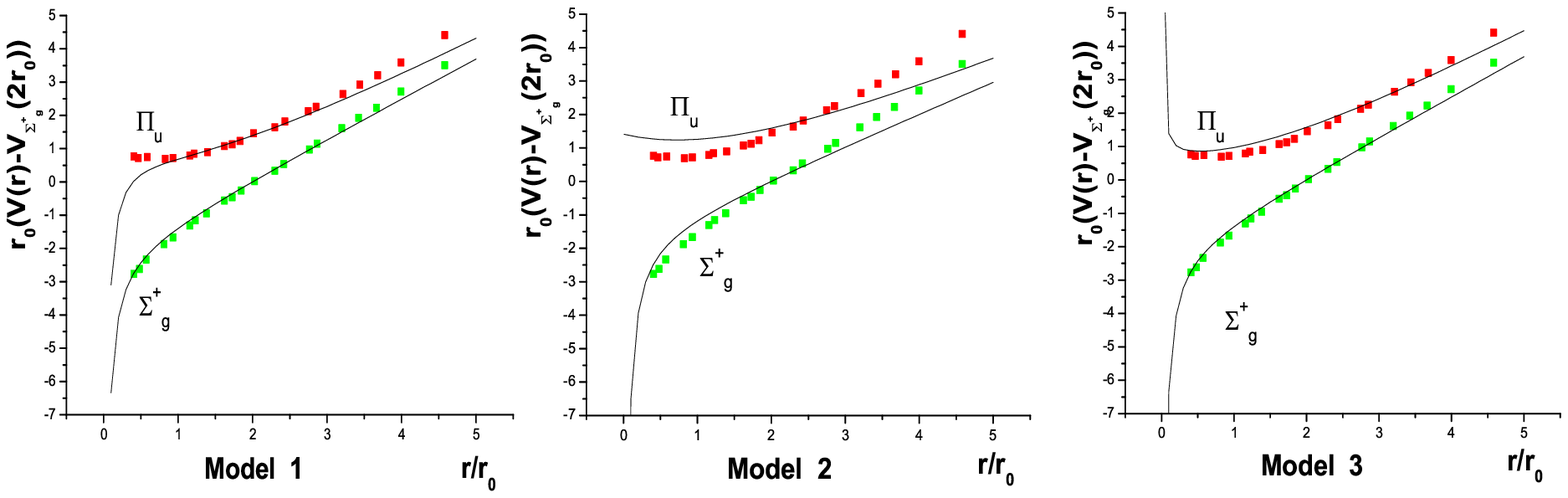}}
\end{tabular}
\end{center}
\caption{}
\end{figure}

\subsection{With the spin-related terms $V_s$}

For the regular quarkonia we adopt the non-relativistic
potential(NR) Eq.(\ref{s5}) \cite{ss}. Since we add a zero-point
energy $V_0$ in the potential which can be seen as another free
parameter (it is the same for both $c\bar c$ and $b\bar b$
quarkonia), we re-fit the spectra of the quarkonia to obtain the
corresponding potential parameters in Eq.(\ref{s5}). We list the
resultant values of the parameters in Table \ref{t3}. In Table
\ref{t4}, we present the fitted spectra of $c\bar c$ and for a
comparison, we also include the results given in Ref.\cite{ss} in
the table.

\begin{table}[!h]
\caption{ potential parameters for $c\bar{c}$}
\begin{center}
\begin{tabular}{|c|c|c|c|c|c|}
  \hline
 $\kappa$ &$b$(GeV$^2$) &$m$(GeV)& $\sigma$(GeV$^2)$&$V_{0}$(GeV)\\\hline
     0.67& 0.16 & 1.78 &1.6  &-0.6\\\hline
\end{tabular}\label{t3}
\end{center}
\end{table}

\begin{table}[!h]
\caption{ Eignvalues for $c\bar{c}$ in GeV}
\begin{center}
\begin{tabular}{|c|c|c|c|c|c|c|}
  \hline
$$ & $J/\psi$ &$\psi(2S)$ &$\psi(3S)$& $\psi(4S)$&$\eta_c(1S)$&$\eta_c(2S)$\\\hline
 Ref\cite{ss}&3.090 & 3.672& 4.072 & 4.406 &2.982 &3.630 \\\hline
   this work& 3.097 & 3.687 & 4.093 & 4.433 & 2.971  & 3.634\\\hline

\end{tabular}\label{t4}
\end{center}
\end{table}

For the $b\bar b$ quarkonia, the corresponding parameters obtained
by fitting data are listed in Table \ref{t5}.

\begin{table}[!h]
\caption{ potential parameters  for $b\bar{b}$}
\begin{center}
\begin{tabular}{|c|c|c|c|c|c|}
  \hline
 $\kappa$ &$b$(GeV$^2$) &$m$(GeV)& $\sigma$(GeV$^2)$)&$V_{0}$(GeV)\\\hline
     0.53&0.16 &5.13  &1.7&-0.60 \\\hline
\end{tabular}\label{t5}
\end{center}
\end{table}

By the parameters we predict $m_{\eta_b}=9.434$ GeV, which is
consistent with that given by \cite{bb}.

Then we turn to the hybrid intermediate states.

For the hybrids, by the observation made in the previous
subsection one can conclude that the third potential (model 3)
better coincides with the lattice results, therefore, in this
subsection when we include the spin-related term to discuss
spin-splitting case,  we only adopt the third potential
Eq.(\ref{h7}). It is reasonable to keep the values of $m_c$, $m_b$
and $\sigma$ to be the same as that we determined for pure $q\bar
q$ quarkonia and we also set $f=1$. Then following our strategy
discussed in previous subsections, we obtain the potential
parameters which are listed in the following table.

\begin{table}[!h]
\caption{ potential parameters for hybrid}
\begin{center}
\begin{tabular}{|c|c|c|c|c|c|}
  \hline
$$ & $\kappa(c\bar{c}g)$&$\kappa(b\bar{b}g)$ &$b$(GeV$^2$) &$V_{0}$(GeV)\\\hline

    the best fitted values &0.54 & 0.40 & 0.24 &-0.80\\\hline

\end{tabular}\label{t6}
\end{center}
\end{table}

The fitted values and some predictions are also listed in Tables
\ref{t7} and \ref{t8}. We obtain
\begin{eqnarray*}\label{s10}
|C_1|^2= 182.12\times10^{-6}, \qquad
\end{eqnarray*}
the mass of hybrids are  4.351GeV, 4.333 GeV for the spin-triplet
and spin-singlet $c\bar c$ in the hybrid and  10.916GeV, 10.913GeV
for the spin-triplet and singlet $b\bar b$ respectively. Because
of including the spin-related term,  the ``ground states" with the
$q\bar q$ (q=b or c) being in different spin structures would be
slightly split.

One can observe that the predicted $\Gamma(\Upsilon(4S)\rightarrow
\Upsilon(3S)\pi\pi)$ and $\Gamma(\psi(3S)\rightarrow
\psi(2S)\pi\pi)$ are slightly smaller than that predicted in the
models without the spin-related term,  the future experiments may
shed some light on it, namely getting better understanding on the
mechanisms which one can describe the hybrid structure better.

We also calculate the transition rate of
$\eta'_c\rightarrow\eta_c+\pi+\pi$, our result is almost triple
that obtained in Ref.\cite{Y1} and it can be tested by the future
experiments.
\begin{table}[!h]
\caption{$\Upsilon$\, transition(in units of keV)}
\begin{center}
\begin{tabular}{|c|c|c|c|c|c|}
  \hline
 decay mode & widths (fit)  \\\hline
   $\Upsilon(2S)\rightarrow \Upsilon(1S)\pi\pi$ &8.73 \\
  $\Upsilon(2S)\rightarrow \Upsilon(1S)\pi\pi$ &1.94 \\
   $\Upsilon(3S)\rightarrow \Upsilon(2S)\pi\pi$ &0.69\\
  $\Upsilon(4S)\rightarrow \Upsilon(1S)\pi\pi$ &4.10\\

  $\Upsilon(4S)\rightarrow \Upsilon(2S)\pi\pi$ &1.88\\
  \hline
\end{tabular}\label{t7}
\end{center}
\end{table}
It is noted that since we minimize $\bar\chi^2$,  the decay widths
that we obtain are different from the central values of the
measured quantities. We list the widths we finally obtained in the
table \ref{t7}.

\begin{table}[!h]
\caption{prediction(in units of keV)}
\begin{center}
\begin{tabular}{|c|c|c|c|c|c|}
  \hline
 decay mode &widths of  predition \\\hline
   $\Upsilon(4S)\rightarrow \Upsilon(3S)\pi\pi$ &0.36 \\
  $\psi(3S)\rightarrow \psi(2S)\pi\pi$ &8.84\\
   $\psi(3S)\rightarrow J/\psi\pi\pi$ &12.38\\
   $\eta_c(2S)\rightarrow\eta_c\pi\pi$&335.66\\
   \hline
\end{tabular}\label{t8}
\end{center}
\end{table}

\section{Our conclusion and discussion}

Search for exotic states which are allowed by the SU(3) quark
model and QCD theory is very important for our understanding of
the basic theory, but so far such states have not been found (or
not firmly identified), thus it becomes an attractive task in high
energy physics. No doubt, direct measurements on such exotic
states would provide definite information on them, however, it
seems that most of the mysterious states mix with mesons and
baryons which have regular quark structures. Since they are hidden
in the mixed states, they are not physical states and do not have
physical masses, and it makes a clear identification of such
exotic states very difficult. In other words, they may only serve
as a component of physical states. Even though, some
phenomenological models, such as the color-flux-tube model, the
bag model and the potential model etc., are believed to properly
describe their properties and determine their ``masses", in fact,
if they mix with the regular mesons or baryons, the resultant
masses are only the diagonal elements of the Hamiltonian matrix.
For example, in the potential model, by solving Schr\"odinger
equation, one obtains the eigen-energy and wave function, he only
gets the element $E_{11}= _{hyb}<\phi|H_{hyb}|\phi>_{hyb}$, where
the subscript ``hyb" denotes the quantities corresponding to
hybrids. Meanwhile, there is $E_{22}=
_{reg}<\phi|H_{reg}|\phi>_{reg}$ corresponding to the regular
quark structure. If the two eigen-states are not far located, they
may mix with each other and provide an extra matrix element to the
hamiltonian matrix, as $E_{12}=E_{21}^*=
_{hyb}<\phi|H_{mix}|\phi>_{reg}$. Unfortunately, there is not a
reliable way to calculate the mixing matrix element. One may
expect to gain definite information about the hybrid states and
maybe starting from there he can further study the mechanism of
the mixing.

The theoretical framework established by Yan and Kuang confirms
that the intermediate states between two pion-emissions in the
transition $\psi(ns)(\Upsilon(ns))\rightarrow
\psi(ms)(\Upsilon(ms))+\pi\pi$, are hybrids which contain a
quark-antiquark pair in color octet, and an extra valence gluon.
Based on the color-flux-tube model, in 80's of last century Isgur
and Paton suggested a potential model for the hybrid, and this
greatly simplifies the discussion about hybrids and may offer an
opportunity to study the regular quarkonium and hybrid in a unique
framework. After their work, several other groups also proposed
modified potentials to make a better description on the hybrid
states. When Yan and Kuang studied the transitions, there were not
many data available, i.e. most of the channels were not measured
yet. Therefore they assumed that $\psi(4.03)$ as the ground state
of charmed hybrids $|c\bar cg>$ and estimated the transition
rates. Thanks to the great achievements of the Babar and Belle
collaborations, many such modes are measured with appreciable
accuracy. Based on the experimental data and the theoretical
framework established by Yan and Kuang, we minimize the
$\bar\chi^2$ to obtain the model parameters in the potential for
hybrid, and with them, we can estimate the masses of the ground
states of hybrids. The theory of the QCD multi-expansion is based
on the assumption that the hadronization of the emitted gluons can
be factorized from the transition of $\Upsilon(ns) (\psi(ns))\to
\Upsilon (ms) (\psi(ms))$. In fact, this factorization may be not
complete if the non-perturbative QCD effects are invloved, namely
the higher twist contribution may somehow violate the
factorization. However, as long as the non-perturbative QCD
effects are not too strong, this approximation should be
acceptable within a certain tolerance range. Moreover, in our
study, the non-factorization effects are partly involved in the
parameter $|C_1|^2$ of Eq. (1), and in our scheme it is also one
of the free parameters  which are fixed by fitting data. Indeed,
it is implicitly assumed that $|C_1|^2$ is universal for all the
processes, and it may cause some error. But it is believed that
since the energy range does not change drastically, the error
should controllable.

In the calculations, we adopt the Cornell potential for the
color-singlet $q\bar q$ (q=b or c) system and the potentials
suggested by Isgur and Paton  (model 1)\cite{Isgur}, by Swanson
and Szczepaniak (model 2) \cite{Swanson} and by Allen $et\, al$
(model 3) \cite{Allen} to deal with the color-octet $q\bar q$
system, we add a spin-related term to the potential for hybrid
(model 3 only) to investigate possible spin-splitting effects. The
numerical results are slightly different when this term is
introduced. The masses of the ground state hybrids are 4.23 GeV
for $|c\bar cg>$ and 10.79 GeV for $|b\bar bg>$ which are
estimated in terms of model 3. When the spin-related term is
included, the results change to  4.351 GeV, 4.333 GeV for the
spin-triplet and spin-singlet $c\bar c$ in the hybrid and 10.916
GeV, 10.913 GeV for the spin-triplet and singlet $b\bar b$
respectively. In other two models, the results are slightly
different. Indeed as aforementioned, a comprehensive comparison of
the results with the lattice values, one may be convinced that the
model 3 may be the best choice at present. All the obtained masses
are different from the physical states measured in experiments,
and it may imply that the hybrids mix with regular mesons.

There are more data in the b-energy range than in charm-energy
region. In fact, when we use the same method to calculate the
transition $\psi(ns)\rightarrow\psi(ms)+\pi\pi$, with n and m
being widely apart (say n=4, m=1 etc.), the theoretical solutions
are not stable and uncertainties are relatively large. It
indicates that there are still some defects in the theory which
would be studied in our future works. Moreover, recently Shen and
Guo \cite{shen} studies the processes in terms of the chiral
perturbation theory and considered the final state interaction to
fit the details of the $\pi\pi$ energy and angular distributions.

The transition of higher excited states of quarkonia into lower
ones (including the ground state) without flavor change but
emitting photon or light mesons is believed to offer rich
information on the hadron structure and governing dynamics,
especially for the heavy quarkonia physics, for example, Brambilla
$et$ $al$.\cite{Brambilla} studied the quarkonium radiative decays
which are realized via electromagnetic interactions.

Our studies indicate that the transitions of
$\psi(ns)(\Upsilon(ns))\rightarrow \psi(ms)(\Upsilon(ms))+\pi\pi$
may provide valuable information about the hybrid structures which
have so far not been identified in experiments yet.

Since we use the method of minimizing $\bar\chi^2$ to achieve all
the parameters in the potential model for hybrids, it certainly
brings up some errors. It is a common method for both
experimentalists and theorists to analyze data and obtain useful
information. Definitely, the more data are available, the more
accurate the results would be. Therefore more data are very
necessary, especially the data on the $\psi$ families which are
one of the research fields of the BES III and CLEOc.

Acknowledgements:

This work is supported by the National Natural Science Foundation
of China (NNSFC), under the contract No. 10475042.\\


\begin{thebibliography}{99}
\bibitem{Gottfried} K. Gottfried, Phys. Rev. Lett. {\bf 40}, 598(1978).

 \bibitem{YK1} Y. Kuang and T.  Yan, Phys. Rev. {\bf D 24}, 2874(1981).

 \bibitem{K2} Y. Kuang , Front. Phys. China {\bf 1}, 19(2006).
  \bibitem{Y1} T. Yan, Phys. Rev. {\bf D 22}, 1652(1980).
 \bibitem{K1} Y.  Kuang, Y. Yi and B. Fu, Phys. Rev. {\bf D 42}, 2300(1990).
 \bibitem{soft} L. Brown and R. Cahn, Phys. Rev. Lett. {\bf 35}, 1(1975).
  \bibitem{BT}W. Buchm$\ddot{\rm u}$ller and H. Tye, Phys. Rev. Lett. {\bf 44}, 850(1980).


\bibitem{PDG} W. Yao $et$ $al$., [Partical Data Group], J. Phys. G
 {\bf 33}, 1(2006).

 \bibitem{4s2s} B. Aubert $et\, al$., [BaBar Collaboration], Phys. Rev. Lett. {\bf 96}, 232001(2006).
   \bibitem{Belle} A. Sokolov $et\, al$., [Belle Collaboration], hep-ex/0611026.

  \bibitem{Isgur} N. Isgur and J. Paton, Phys. Rev. {\bf D 31}, 2910(1985).
\bibitem{Swanson} E. Swanson and A. Szczepaniak, Phys. Rev. {\bf D 59}, 014035(1999).
\bibitem{Allen} T.  Allen, M. Olsson and S. Veseli,  Phys. Lett. {\bf B 434},
110(1998).


\bibitem{chi} C. Chiang, M. Gronau, J. Rosner and D. Suprun, Phys. Rev. {\bf D 70}, 034020(2004).
 \bibitem{cornell} E. Eichten, K. Gottfried, T. Kinashita, K. Lane and T. Yan
 , Phys. Rev. {\bf D 17}, 3090 (1978); $ibid$, {\bf D 21}, 203(1980).
% \bibitem{3exp} K. Abe $et\, al$., [Belle Collaboration], Phys. Rev. {\bf D 70}, 071102 (2004);
% D. Asner $et\, al$., [Cleo Collaboration], Phys. Rev. Lett. {\bf 92}, 142001, (2004);
%B. Aubert $et\, al$., [BaBar Collaboration], Phys. Rev. Lett. {\bf 92}, 142002, (2004).

 \bibitem{ss} T. Barnes, S. Godfrey and E. Swanson , Phys. Rev. {\bf D 72}, 054026(2005) .
\bibitem{Buisseret} F. Buisseret and V. Mathieu, Eur. Phys. J. {\bf A 29}, 343(2006).
\bibitem{Szczepaniak}  A. Szczepaniak and P. Krupinski, Phys. Rev. {\bf D 73},
116002(2006); P. Guo, A. Szczepaniak G. Galata, A. Vassallo and E.
Santopina, arXiv:0707.3156 [hep-ph].
\bibitem{Bali}  G. Bali and A. Pineda, Phys. Rev. {\bf D 69}, 094001(2004).
\bibitem{lattice} K. Juge, J. Kuti and C. Morningstar, Nucl. Phys. (Proc. Suppl.) {\bf B 63}, 326(1998);
 Phys. Rev. Lett. {\bf 90}, 161601 (2003).

 \bibitem{bb} G. Hao, Y. Jia, C. Qiao, and P. Sun, Phys. Rev. {\bf D 75}, 035010(2007);
 D. Ebert, R.  Faustov and V.  Galkin, Phys. Rev. {\bf D 67}, 014027(2003);
  S. Recksiegel and Y. Sumino, Phys. Lett. {\bf B 578}, 369(2004);
B.  Kniehl, A.  Penin, A. Pineda, V.  Smirnov and M. Steinhauser,
Phys. Rev. Lett. {\bf 92}, 242001(2004); A. Gray, I. Allison, C.
Davies, E. Gulez, G.  Lepage, J. Shigemitsu and M. Wingate, Phys.
Rev. {\bf D 72}, 094507(2005).

\bibitem{shen} F. Guo, P. Shen, H. Chiang and R. Ping, Nucl. Phys. {\bf {A761}},
269(2005); F. Guo, P. Shen and H. Chiang, Phys. Rev. {\bf {D74}},
014011(2006); F. Guo, P. Shen, H. Chiang and R. Ping,
hep-ph/0601120.
\bibitem{Brambilla} N. Brambilla, Y. Jia and A. Vairo, Phys. Rev. {\bf D 73},
054005(2006); N. Brambilla $et\, al$., hep-ph/0412158.


\end{thebibliography}
\end{document}